# scientific reports

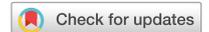

### OPEN

# Glycine amino acid transformation under impacts by small solar system bodies, simulated via high-pressure torsion method

Kaveh Edalati[1]✉, Ikuo Taniguchi[1], Ricardo Floriano[2] & Augusto Ducati Luchessi[2,3]

Impacts by small solar system bodies (meteoroids, asteroids, comets and transitional objects) are characterized by a combination of energy dynamics and chemical modification on both terrestrial and small solar system bodies. In this context, the discovery of glycine amino acid in meteorites and comets has led to a hypothesis that impacts by astronomical bodies could contribute to delivery and polymerization of amino acids in the early Earth to generate proteins as essential molecules for life. Besides the possibility of abiotic polymerization of glycine, its decomposition by impacts could generate reactive groups to form other essential organic biomolecules. In this study, the high-pressure torsion (HPT) method, as a new platform for simulation of impacts by small solar system bodies, was applied to glycine. In comparison with high-pressure shock experiments, the HPT method simultaneously introduces high pressure and deformation strain. It was found that glycine was not polymerized in the experimental condition assayed, but partially decomposed to ethanol under pressures of 1 and 6 GPa and shear strains of < 120 m/m. The detection of ethanol implies the inherent availability of remaining nitrogen-containing groups, which can incorporate to the formation of other organic molecules at the impact site. In addition, this finding highlights a possibility of the origin of ethanol previously detected in comets.

RNA (ribonucleic acid) and protein are two essential polymer molecules for life, but it has not been clear how these biomolecules were formed about four billion years ago on the Earth from nucleosides (such as adenosine monophosphate) and amino acid monomers, respectively[1]. The discovery of molecules of amino acids such as glycine ($C_2H_5NO_2$) in the Murchison meteorite (fell in Australia in 1969) suggested that impacts by small solar system bodies (such as meteoroids, asteroids, comets, and transitional objects) could have brought the important organic molecules for life into the Earth[2]. Recent examination of various types of carbonaceous chondrite meteorites including Tagish Lake meteorite (fell in Canada in 2000) also confirmed the presence of amino acids, which could survive the atmospheric entry, impact effect and aqueous alternation[3]. Moreover, theoretical calculations suggested that high pressure, high temperature and high strain rate during astronomical impact events could form protein chains from amino acids[4]. So far, there has not been strong experimental evidence for the formation of protein from amino acids by impact simulation via the high-pressure shock experiments[5–8].

For impacts by small solar system bodies, pressure, temperature, and strain rate at the impact point are high, but they are reduced with increasing the distance from the collision point[9]. Moreover, the deformation strain occurs at different distances away from the impact point, as shown schematically in Fig. 1a[10]. While the shock experiments are appropriate to simulate the high-pressure conditions at the impact point, they are not effective to simulate the conditions at some distances from the collision point where the deformation strain is high, and the pressure and temperature are rather low[11]. So far, there have been few attempts to consider the simultaneous effect of high pressure and high strain on polymerization or stability of amino acids.

The high-pressure torsion (HPT) method can be an appropriate method to simulate astronomical impact events because in addition to pressure, temperature and strain rate, strain can be simultaneously controlled in this method[12]. A disc-shaped sample is compressed between two anvils under high pressure and strain is induced by rotating one of the anvils with respect to the other one in this method, as shown in Fig. 1b[13]. In the last century, scientists from the USA[14,15] and the USSR[16,17] applied the HPT method to some organic compounds and reported enhanced polymerization or depolymerization; however, these interesting studies were virtually overlooked later as discussed in a review paper[13]. The HPT method is currently used as a severe plastic deformation technique[18,19] for controlling the crystal size[12,20] and phase transformation[21,22] in inorganic materials. The method is considered quite effective to connect atoms of various pure elements to generate larger molecules of metallic alloys[23] and multi-component ceramics[24].

[1]WPI International Institute for Carbon-Neutral Energy Research (WPI-I2CNER), Kyushu University, Fukuoka, Japan. [2]School of Applied Sciences, University of Campinas (UNICAMP), Limeira, São Paulo, Brazil. [3]Institute of Biosciences, São Paulo State University (UNESP), Rio Claro, São Paulo, Brazil. ✉email: kaveh.edalati@kyudai.jp

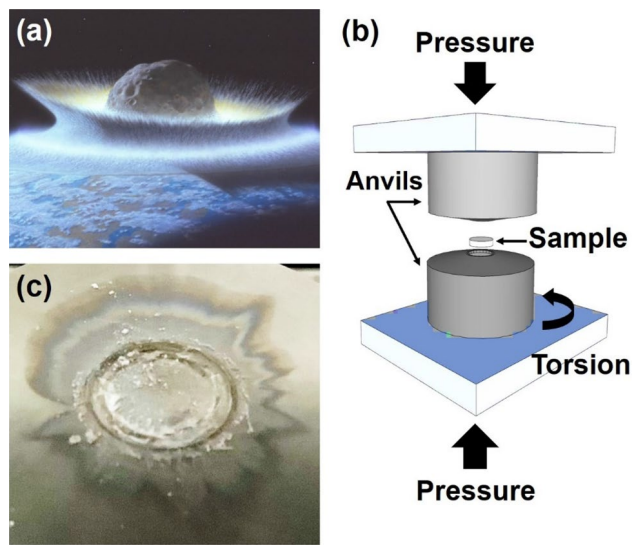

**Figure 1.** (**a**) Illustration of impacts by meteoroids, asteroids, comets and transitional objects on the Earth[10]; (**b**) schematic of high-pressure torsion method and its anvils[13]; and (**c**) surface of high-pressure torsion anvil after a small explosion caused by processing glycine under 6 GPa.

In this study, we apply the HPT method for the first time to glycine amino acid to have an insight into the survival of amino acids or the formation of protein by astronomical impact events. The findings of this study are of significance for astrobiology because the degree of survival of organic materials during impacts by small solar system bodies is an important issue in understanding the delivery of necessary molecules for life onto the Earth by comets and other small solar system bodies[25,26]. Moreover, the formation mechanism of proteins from amino acids in the early Earth conditions is still a key issue in understanding the origin of life[1,8].

## Methods

Glycine powder with a chemical composition of $C_2H_5NO_2$, a purity level of >99%, a density of 1.2 kg/m$^3$, and a crystal structure of monoclinic (α phase with a P21/n space group) was selected for this study. The powder was compacted under 0.3 MPa to discs with 10 mm diameter and 0.8 mm thickness (almost 80 mg mass of each disc). The disc samples were processed between two cylindrical HPT anvils with 50 mm diameter and 50 mm height made from the WC—11 wt% Co composite. The surfaces of anvils had a 5° slope from the center to edge with a flat-bottomed hole with 10 mm diameter and 0.25 mm height at the center. The compacted discs were processed under pressures of $P = 1$ and 6 GPa at ambient temperature ($T = 300$ K). Although earlier studies suggested that the pressure can be in the range of 5–25 GPa[25] or typically below 30 GPa[5] during impact by extraterrestrial objects, the current study considers that lower pressures such as 1 GPa can be naturally generated with increasing the distance from the collision point due to the distribution of normal pressure vursus distance[19]. Shear strain, $\gamma$ ($\gamma = 2\pi r N/h$, $r$: distance from disc center, $N$: anvil rotations, $h$: sample thickness)[13], was introduced by rotating the lower HPT anvil with respect to the upper one with a rotation speed of 1 rpm (turn per minute) for $N = 3$ rotations. The shear strain was theoretically zero at the disc center and increased to 120 m/m at the disc edge. The sample processed under 6 GPa experienced a miniature explosion with the start of rotation which introduced a significant shock to the sample. Figure 1c shows the appearance of anvil surface after processing glycine under 6 GPa. The initial glycine powder and HPT-processed samples were examined by different techniques, as described below.

Crystal structure was examined by X-ray diffraction (XRD) method using the Cu $K\alpha$ radiation with a wavelength of 0.15406 nm (45 kV and 200 mA) and by Raman spectroscopy using a 532-nm laser source. XRD profiles were achieved using a complete disc, and thus, they represent the crystal structure from the center to the edge of disc, i.e., from shear strain of zero to 120 m/m. Raman spectra were recorded at highly strained region at 4 mm away from the disc center, although no clear difference was detected between the Raman spectra at the disc center and edge.

The polymerization or decomposition of samples was examined by nuclear magnetic resonance (NMR) spectroscopy. About 20 mg of glycine samples were dissolved in 0.6 mL of $D_2O$ and placed into an NMR tube (inner diameter: 5 mm). $C_6D_6$ in a glass capillary was also placed in the NMR tube as an internal standard. The $^1$H and $^{13}$C NMR spectra were recorded at 600 and 150 MHz, respectively.

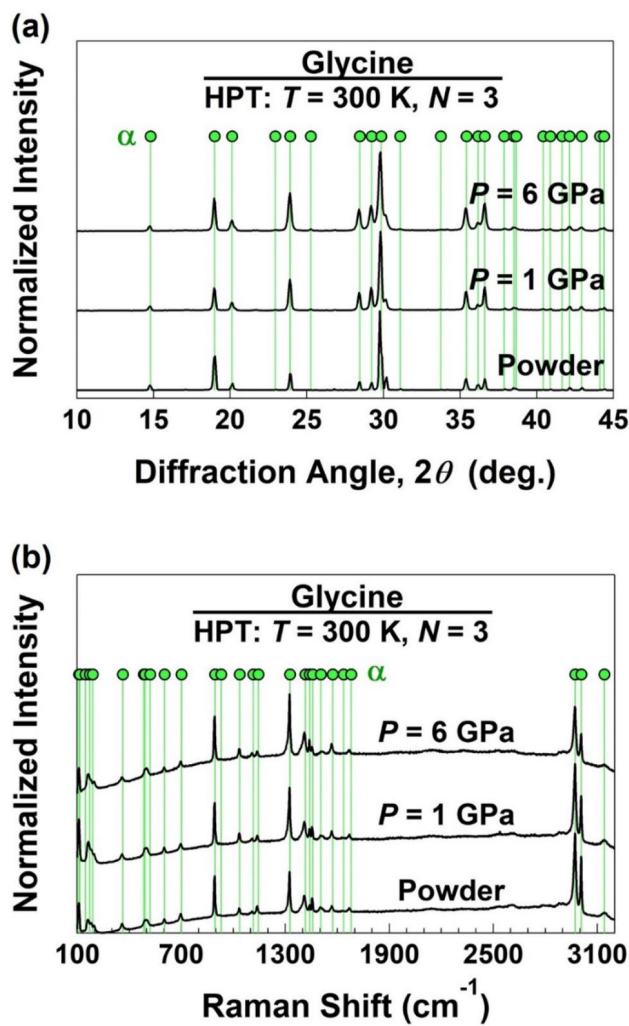

**Figure 2.** Absence of phase transformation in glycine after high-pressure torsion processing. (**a**) XRD profiles and (**b**) Raman spectra of glycine before and after processing under 1 and 6 GPa.

### Results

Examination of crystal structures by XRD analysis is shown in Fig. 2a. A comparison between the XRD profile of the initial powder and the reference data (PDF Card No.: 00-032-1702) confirms that the initial powder contains only α-glycine. Crystal structure remains the same after HPT processing and no transformations occur to the β, γ or δ phases or other high-pressure polymorphs of glycine[27,28]. Examination of crystal structure by Raman spectroscopy, as show in Fig. 2b, also confirms that the initial powder and the samples processed by HPT under 1 and 6 GPa have similar spectra. A comparison between these spectra and the reported wavenumbers in the literature[29,30] confirms that the three samples contain pure α-glycine. Taken altogether, Fig. 2 suggests that no irreversible phase transformations occur in glycine by HPT processing, and this is consistent with the reported stability of α-glycine at room temperature and under pressures up to 50 GPa[31].

The polymerization of glycine was examined by (a) $^1$H and (b) $^{13}$C NMR spectroscopy, as show in Fig. 3. For the initial powder, only the peaks corresponding to pure glycine appear in good agreement with the reference data[32]. After HPT processing under 1 and 6 GPa, the peaks originated from glycine peaks are found, which is in good agreement with the XRD analysis and Raman spectroscopy. No broadening or shift of the glycine peaks suggests the absence of polymerization. Besides the glycine and reference $C_6D_6$ peaks, peaks at 3.65 and 1.17 ppm in $^1$H NMR and 17.5 and 58.0 ppm in $^{13}$C NMR can be assigned to ethanol (ethyl alcohol, $C_2H_6O$) after the HPT processing[25]. The formation of ethanol suggests that glycine is decomposed and not polymerized by the HPT processing conditions. The nitrogen-containing or amino-group-based decomposition product could not be traced by NMR technique suggesting that they evaporate/react fast and move into the air atmosphere, while the evolved ethanol remains for sufficient time in the sample during the NMR spectroscopy measurements.

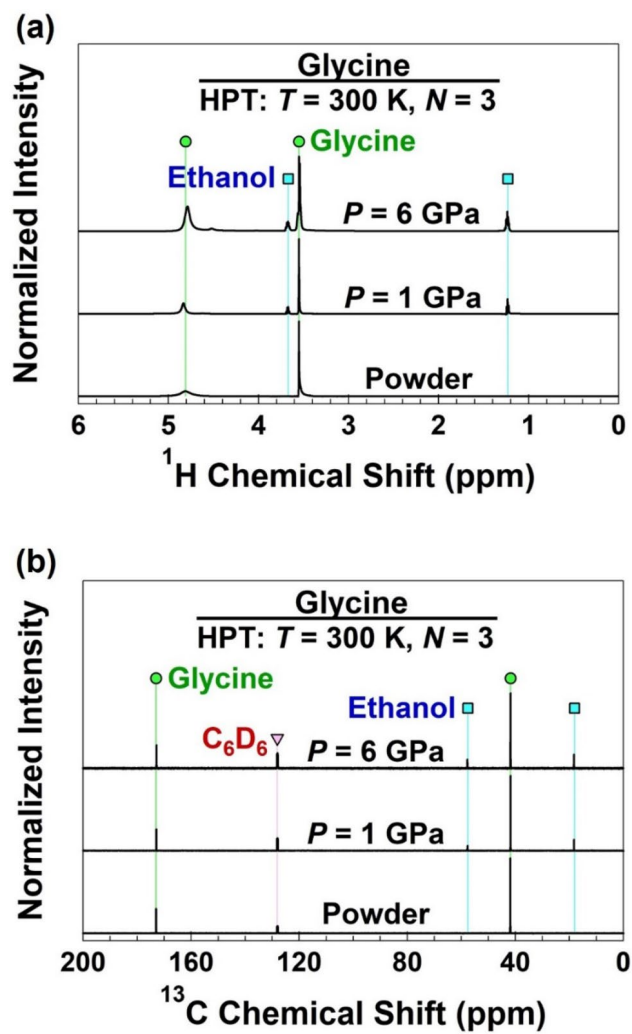

**Figure 3.** Absence of polymerization and occurrence of transformation to ethanol in glycine by high-pressure torsion processing. NMR (**a**) $^1$H and (**b**) $^{13}$C spectra of glycine before and after processing under 1 and 6 GPa.

## Discussion

The current results on the first application of the HPT method to glycine amino acid does not provide any evidence for the polymerization and formation of protein, but large amounts of glycine can survive under simultaneous application of high pressure and shear strain. Since glycine is one of the stable amino acids with a low heat of formation, the absence of polymerization is not surprising. Earlier shock experiments by Peterson et al. showed that polymerization of amino acids does not occur under pressure range of 3.5 to 32 GPa[5]. Blank et al. also showed that significant amounts of amino acids can survive under high pressures (5.1–21 GPa) and temperatures (412–870 K)[6]. However, the formation of ethanol from glycine by additional shear strain effect, which was observed for the first time in this study, is rather unusual because ethanol has a higher heat of formation compared to glycine. The formation of ethanol should be due to the significant energy induced during the HPT process[12,13], which can change the thermodynamic stabilities under high pressure and strain[21,22]. Earlier studies on HPT processing confirmed that such thermodynamic changes can move to different directions[19] such as decomposition of large molecules[22] or combination of small atoms to form larger molecules[22–24] even when the small atoms are thermodynamically immiscible[33]. For organic materials, classic studies also showed that decomposition or polymerization can occur under high pressure and high strain, depending on the material compositions, suggesting that the thermodynamic stabilities can be influenced in different directions in organic materials as well[14–17].

Here, it should be noted that it is believed that alcohols including ethanol could contribute to the evolution of organic compounds in the early Earth[34] and there are also recent reports on the discovery of ethanol in comets[35]; however, the direction of such contributions still needs to be investigated. Since the presence of glycine was confirmed at least in two comets 81P/Wild-2[7] and 67P/Churyumov-Gerasimenko[36], the current experiments suggest that glycine may be a natural source of huge amounts of ethanol discovered in some comets[35]. It should be noted that comets can experience shear strain under tidal forces of planets and stars. While some big comets such as C/2014 Q2 (Lovejoy), in which alcohol was discovered, can survive these tidal forces[35], some other comets experience disruption by these forces such as Kreutz comets disrupted by sun[37] or comet D/1993 F2

Shoemaker-Levy 9 disrupted by Jupiter[38]. Moreover, the gravitational forces of planets and stars can lead to the impact of these sheared comets by other astronomical bodies, providing sufficient pressure for transformation of amino acids to alcohol particularly in the presence of iced water, aqueous alteration and mineral catalysts[39].

Despite the formation of ethanol from glycine by HPT processing, it is still hard to clarify the pathway for this transformation because the HPT experiments are conducted under ambient atmosphere and thus the gas phases cannot be analyzed. Conversion of glycine have been widely studied under different conditions such as high pressure[8], thermal heating[40], ionizing radiation[41], protonated condition[42], high-pressure/temperature water[43], and electrochemical condition[44]. None of these studies reported the formation of ethanol from glycine, although electrochemical formation of methanol and propanol alcohols was mentioned in a publication[45]. Although tracking the transformation pathway of glycine is experimentally difficult due to the short lifetime of radicals[46] as well as due to the formation of various gaseous species (such as $H_2$, $H_2O$, $CO$, $CO_2$, $N_2$ and $NH_3$)[41–44], computational studies suggested some possible pathways and mechanisms for some of the transformations[47,48]. By considering the available information in the literature and by considering that the HPT process was conducted under the air atmosphere, three possible transformation pathways can be suggested: (i) decomposition of glycine by severe energy induced by the HPT method, (ii) reaction of glycine with water from humidity, and (iii) oxidation of glycine with air. Since the XRD, Raman and NMR analyses detected only glycine and ethanol, it is assumed that other products are mainly in the gas form. The explosion occurred during the HPT process also confirms that reactive gases should have formed. The absence of an ammonia smell during the HPT process suggests that $NH_3$ with appreciable quantities might not have formed, although $NH_3$ formation by glycine decomposition was reported in many publications[41–45,49]. The absence of $NH_3$ suggests that high energy introduced by HPT not only promotes the elimination reaction to produce amino group by breaking the C-N bonds, but also breaks the N–H bonds, which can subsequently lead to the formation of nitrogen and hydrogen gas phases. Another possibility is the reaction of amino group with oxygen and formation of hydrogen gas and NO, which can subsequently transform to $NO_2$ in the presence of oxygen. Moreover, as suggested in some computational studies[47,48], decarboxylation associated with proton transfer, is expected to lead to the formation of CO or $CO_2$. Here, three possible examples for (1) decomposition, (2) reaction with water and (3) oxidation of glycine are suggested.

$$2C_2H_5NO_2 \rightarrow C_2H_6O + H_2 + H_2O + 2CO + N_2 \quad (1)$$

$$2C_2H_5NO_2 + H_2O \rightarrow C_2H_6O + 3H_2 + 2CO + N_2 \quad (2)$$

$$2C_2H_5NO_2 + 7/2O_2 \rightarrow C_2H_6O + 2H_2O + CO_2 + NO_2 \quad (3)$$

These speculative transformation pathways show the possibility for the glycine-to-ethanol transformation by external energy, but future studies are required to clarify the exact chemical reaction path and its underlying mechanism.

Although current HPT experiments did not provide any positive support for the formation of protein from glycine by astronomical impact events, the formation of protein from other amino acids by HPT processing needs to be investigated in the future. The stability of significant amounts of glycine after the HPT process indicates that comet impacts could basically deliver significant amounts of biomolecules into the early Earth, as suggested in some other studies by high-pressure shock experiments as well[25,26]. As discussed in the previous paragraph, the conversion of glycine to ethanol implies the release of nitrogen-containing molecules and some reactive gases into the system, which are inevitably available to react with other molecules to produce other organic compounds on the Earth. One main difference between this study and real impacts by small solar system bodies is that the current HPT experiments were conducted on high-purity glycine; while amino acids in the early Earth condition or in comets co-presented with some minerals, chemicals, and water[1,2,8,34]. The presence of water and aqueous alternation in carbonaceous chondrite meteorites can affect the formation of organic compounds[39,50]. Moreover, the reactive minerals in meteorites can act as catalysts to form complex organic compounds including amino acids and alcohols particularly in the presence of hot water[51,52]. Therefore, although the current results introduce a new experimental path to simulate the effect of impacts by comets or other small solar system bodies on organic molecules, future HPT experiments on different amino acids and adenosine monophosphate with different purity levels and different additives in both dry and humid conditions are required. Moreover, theoretical calculations like the ones reported earlier[4] can be useful to find the energetically favorable reaction direction in HPT processing to produce protein and RNA.

## Conclusions
To have an insight into the effect of impacts by small solar system bodies (e.g., meteoroids, asteroids, comets, and transitional objects) on the stability of amino acids or the formation of proteins, the simultaneous effect of high pressure and shear strain on polymerization of glycine amino acid was investigated using the high-pressure torsion method. Glycine was not polymerized, but it was partially decomposed and converted to ethanol by processing. The detection of ethanol not only indicates a natural source of ethanol already detected in comets, but also suggests the formation of other decomposition products which could contribute to the formation of other organic biomolecules on the Earth billions of years ago. Although this study did not provide any evidence for the formation of proteins, it introduced the high-pressure torsion method as a new research tool for simulating the effect of impacts by small solar system bodies on the behavior of organic materials in the early Earth condition.

### Acknowledgements
This work is supported in part by WPI-I2CNER, Japan, and in part by Grants-in-Aid for Scientific Research on Innovative Areas from the MEXT, Japan (19H05176 and 21H00150).

### Author contributions
K.E., I.T., R.F., A.D.L.: conceptualization, methodology, validation, writing paper.

### Competing interests
The authors declare no competing interests.

### Additional information
**Correspondence** and requests for materials should be addressed to K.E.

**Reprints and permissions information** is available at www.nature.com/reprints.

**Publisher's note**  Springer Nature remains neutral with regard to jurisdictional claims in published maps and institutional affiliations.

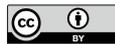